%% file: bare_jrnl.tex
\documentclass[10pt,conference]{IEEEtran}
\usepackage[letterpaper, left=1.02in, right=1.02in, bottom=1in, top=0.75in]{geometry}

\newcommand{\subparagraph}{}
\usepackage[compact]{titlesec}

\usepackage{graphicx}
\graphicspath{ {images/} }
\usepackage{amsmath}
\usepackage{amsfonts}
\usepackage{amssymb}
\usepackage{color}
\usepackage{comment}
\usepackage{subcaption}
\usepackage{multirow}
\usepackage{graphicx}
\usepackage{float}
\usepackage{booktabs}
\usepackage{cleveref}
\usepackage[noadjust]{cite}

\titlespacing{\subsubsection}{0pt}{*0}{*0}

\begin{document}
%
\title{Destruction of Image Steganography \\ using Generative Adversarial Networks }
%
%
%

\author{\IEEEauthorblockN{Isaac Corley\IEEEauthorrefmark{1}, Jonathan Lwowski\IEEEauthorrefmark{2}, Justin Hoffman\IEEEauthorrefmark{3}}
\IEEEauthorblockA{Booz Allen Hamilton \\
San Antonio, Texas \\
Email: \IEEEauthorrefmark{1}corley\_isaac@bah.com,
\IEEEauthorrefmark{2}lwowski\_jonathan@bah.com,
\IEEEauthorrefmark{3}hoffman\_justin@bah.com}}

\newcommand\blankfootnote[1]{%
  \let\thefootnote\relax\footnotetext{#1}%
  \let\thefootnote\svthefootnote%
}

\maketitle

\blankfootnote{Patent Applied for in the United States}

\begin{abstract}
\input{sections/abstract.tex}
\end{abstract}

\begin{IEEEkeywords}
Steganalysis, Deep Learning, Machine Learning, Neural Networks, Steganography, Image Steganalysis, Information Hiding
\end{IEEEkeywords}

%
\IEEEpeerreviewmaketitle

\section{Introduction}
\label{intro}
\input{sections/intro.tex}

\section{Background}
\label{background}
\input{sections/background.tex}

\section{Data}
\label{dataset}
\input{sections/dataset.tex}

\section{Deep Digital Steganography Purification}
\label{dnn}
\input{sections/dnn.tex}

\section{Results}
\label{results}
\input{sections/results.tex}

\section{Conclusion}
\label{conclusion}
\input{sections/conclusion.tex}

\section{Acknowledgements}
\input{sections/acknowledgements.tex}


%





\ifCLASSOPTIONcaptionsoff
  \newpage
\fi



%


\bibliographystyle{IEEEtran}
\bibliography{bibliography}

\end{document}

%% file: sections/abstract.tex
Digital image steganalysis, or the detection of image steganography, has been studied in depth for years and is driven by Advanced Persistent Threat (APT) groups', such as APT37 Reaper, utilization of steganographic techniques to transmit additional malware to perform further post-exploitation activity on a compromised host. However, many steganalysis algorithms are constrained to work with only a subset of all possible images in the wild or are known to produce a high false positive rate. This results in blocking any suspected image being an unreasonable policy. A more feasible policy is to filter suspicious images prior to reception by the host machine. However, how does one optimally filter specifically to obfuscate or remove image steganography while avoiding degradation of visual image quality in the case that detection of the image was a false positive? We propose the Deep Digital Steganography Purifier (DDSP), a Generative Adversarial Network (GAN) which is optimized to destroy steganographic content without compromising the perceptual quality of the original image. As verified by experimental results, our model is capable of providing a high rate of destruction of steganographic image content while maintaining a high visual quality in comparison to other state-of-the-art filtering methods. Additionally, we test the transfer learning capability of generalizing to to obfuscate real malware payloads embedded into different image file formats and types using an unseen steganographic algorithm and prove that our model can in fact be deployed to provide adequate results.

%% file: sections/intro.tex
Steganography is the usage of an algorithm to embed hidden data into files such that during the transferral of the file only the sender and the intended recipient are aware of the existence of the hidden payload \cite{cox2007digital}. In modern day applications, adversaries and Advanced Persistent Threat (APT) groups, such as APT37 Reaper \cite{apt37}, commonly utilize these algorithms to hide the transmission of shellcode or scripts to a compromised system. Once the file is received, the adversary then extracts the malicious payload and executes it to perform further post-exploitation activity on the target machine. Due to the highly undetectable nature of the current state-of-the-art steganography algorithms, adversaries are able to evade defensive tools such as Intrusion Detection Systems (IDS) and/or Antivirus (AV) software which utilize heuristic and rule-based techniques for detection of malicious activity.

Modern steganalysis methods, or techniques developed to specifically detect steganographic content, utilize analytic or statistical algorithms to detect traditional steganography algorithms, such as Least Significant Bit (LSB) steganography \cite{bender1996techniques}. However, these methods struggle to detect advanced steganography algorithms, which embed data using unique patterns based on the content of each individual file of interest. This results in high false positive rates when detecting steganography and poor performance when deployed, effectively making it unrealistic to perform preventative measures such as blocking particular images or traffic from being transmitted within the network. Furthermore, image steganalysis techniques are typically only capable of detecting a small subset of all possible images, limiting them to only detect images of a specific size, color space, or file format.

To address these issues, the usage of steganography destruction techniques can be used to provide a more feasible policy to handling potential false alarms. Instead of obstructing the transmission of suspicious images, filtering said images to remove steganographic content effectively making it unusable by potential adversaries provides a simpler solution. However, traditional and unintelligent steganographic filtering techniques result in the additional issue of degrading image quality. 

In this paper, we propose an intelligent image steganography destruction model which we term Deep Digital Steganography Purifier (DDSP), which utilizes a Generative Adversarial Network (GAN) \cite{goodfellow2014generative} trained to remove steganographic content from images while maintaining high perceptual quality. To the best of our knowledge, our DDSP model removes the greatest amount of steganographic content from images while maintaining the highest visual image quality in comparison to other state-of-the-art image steganography destruction methods detailed in Section \ref{Prior Work}. 

The rest of the paper is organized as follows. Section \ref{background} will discuss the prior work in the image steganography purification domain, along with the background information on GANs. The dataset used for evaluating our image purification model will be presented in Section \ref{dataset}, followed by a detailed description of DDSP in Section \ref{dnn}. The experimental results will be discussed and analyzed in Section \ref{results}. Finally the conclusion and future works will be presented in Section \ref{conclusion}.

%% file: sections/background.tex
\subsection{Prior Work} \label{Prior Work}

Attacks on steganographic systems \cite{westfeld1999attacks} have been an active topic of research with differing objectives of either completely removing hidden steganographic content or slightly obfuscating the steganographic data to be unusable while avoiding significant degradation of file quality. Within the same realm of machine learning based steganographic attacks, the PixelSteganalysis \cite{jung2019pixelsteganalysis} method utilizes an architecture based on PixelCNN++ \cite{salimans2017pixelcnn++} to build pixel and edge distributions for each image which are then manually removed from the suspected image. The PixelSteganalysis method is experimented against three steganographic algorithms, Deep Steganography \cite{baluja2017hiding} a Deep Neural Network (DNN) based algorithm, Invisible Steganography GAN (ISGAN) \cite{zhang2019invisible} a GAN based algorithm, and the Least Significant Bit (LSB) algorithm. The majority of other steganographic destruction techniques are based on non-machine learning based methods which utilize various digital filters or wavelet transforms \cite{amritha2019anti, amritha2016removal, ameen2013optimal, sharp2013novel, al2006destroying, shrestha2011general, amritha2016active, smith2007denoising}. While these approaches may appear simple to implement and do not require training on a dataset, they do not specifically remove artifacts and patterns left by steganographic algorithms. Instead these techniques look to filter out high frequency content, which can result in image quality degradation due to perceptual quality not being prioritized.

\subsection{Super Resolution GAN}
\label{sec:srgan}
The application of GANs for removing steganographic content while maintaining high visual quality draws inspiration from the application of GANs to the task of single image super resolution (SISR). Ledig et. al's research \cite{ledig2017photo} utilized the GAN framework to optimize a ResNet \cite{he2016deep} to increase the resolution of low resolution images to be as visually similar as possible to their high resolution counterparts. Their research additionally detailed that using the GAN framework as opposed solely to a pixel-wise distance loss function, e.g. mean squared error (MSE), results in additional high frequency texture detail to be recreated. They note that using the MSE loss function alone results in images that appear perceptually smooth due to the averaging nature of their objective function. Instead, the authors optimize a multi-objective loss function which is composed of a weighted sum of the pixel-wise MSE as a content loss and the adversarial loss produced by the GAN framework to reconstruct the low and high frequency detail, respectively.

%% file: sections/dataset.tex
To train our proposed model, we utilized the BOSSBase dataset \cite{bossbase} because it is used as a benchmark dataset for numerous steganalysis papers. The dataset contains 10,000 grayscale images of size 512x512 of the portable gray map (PGM) format. The images were preprocessed prior to training by converting to the JPEG format with a quality factor of 95\% and resizing to 256x256 dimensions. As seen in Figure \ref{fig:dataset}, 4 different steganography algorithms are used to embed payloads into the 10,000 images. These algorithms (HUGO \cite{hugo}, HILL \cite{hill}, S-UNIWARD \cite{suniward}, and WOW \cite{wow}) are a few state-of-the-art steganography algorithms which are difficult to detect even by modern steganalysis algorithms. These algorithms are open source and made available by the Digital Data Embedding Laboratory of Binghamton University  \cite{stego_code}. For each of these steganography algorithms, 5 different embedding rates, 10\%, 20\%, 30\%, 40\%, and 50\%, were used with 10\% being the most difficult to detect, and 50\% being the easiest to detect. This process created a total of 210,000 images consisting of 200,000 steganographic images, and 10,000 cover images. The BOSSBase dataset was then split into train and test sets using a train/test split criterion of 75\%/25\%.

\input{figures/dataset_arch.tex}

\input{figures/ddsp_arch.tex}
\input{figures/autoencoder_arch.tex}
\input{figures/encoder_arch.tex}

%% file: figures/dataset_arch.tex
\begin{figure}[h!]
\centering
  \includegraphics[width=0.95\linewidth]{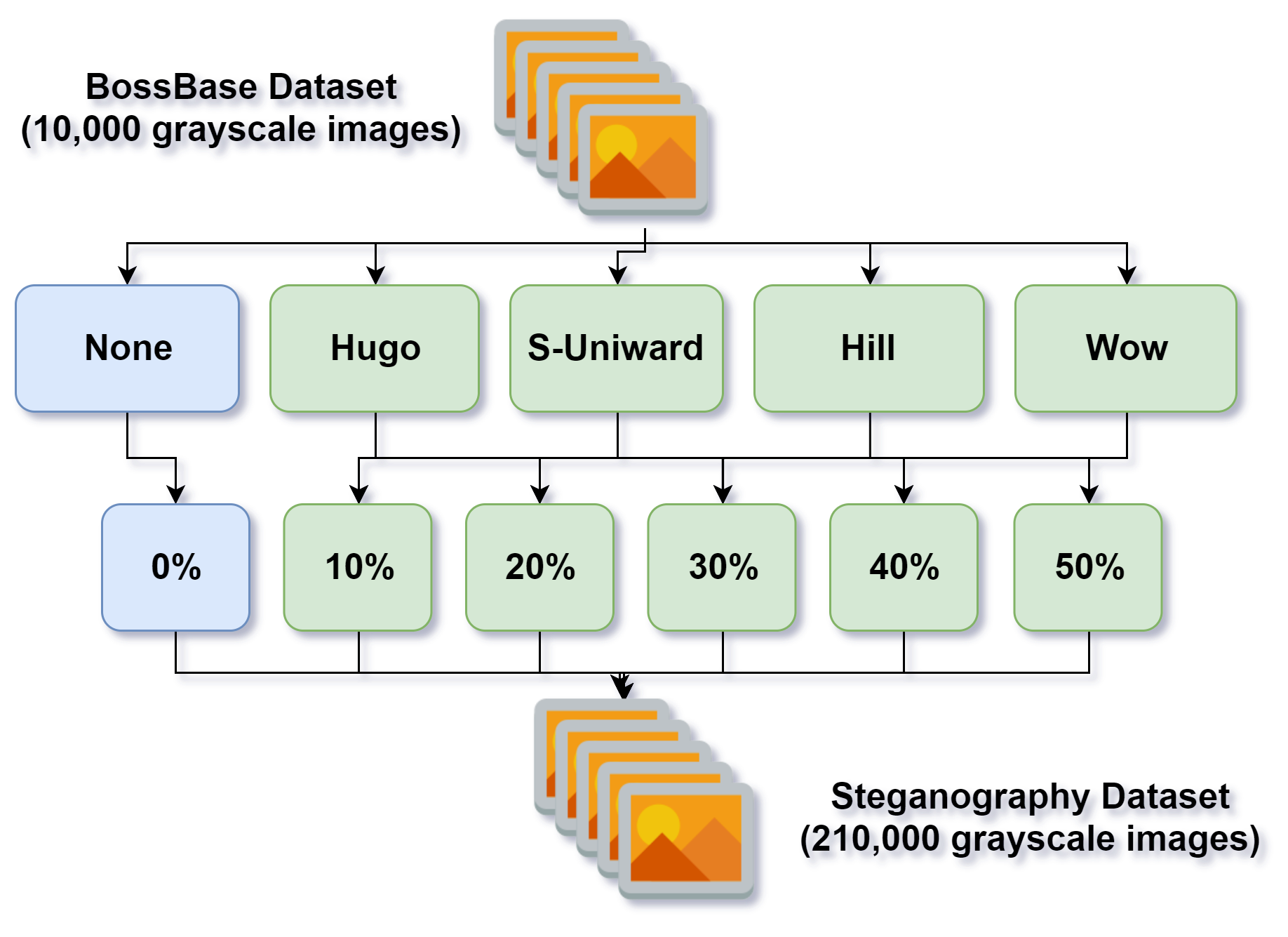}
  \caption{Steganography dataset creation process}
  \label{fig:dataset}
\end{figure}

%% file: figures/ddsp_arch.tex
\begin{figure}[!t]
  \centering
  \includegraphics[width=0.99\linewidth]{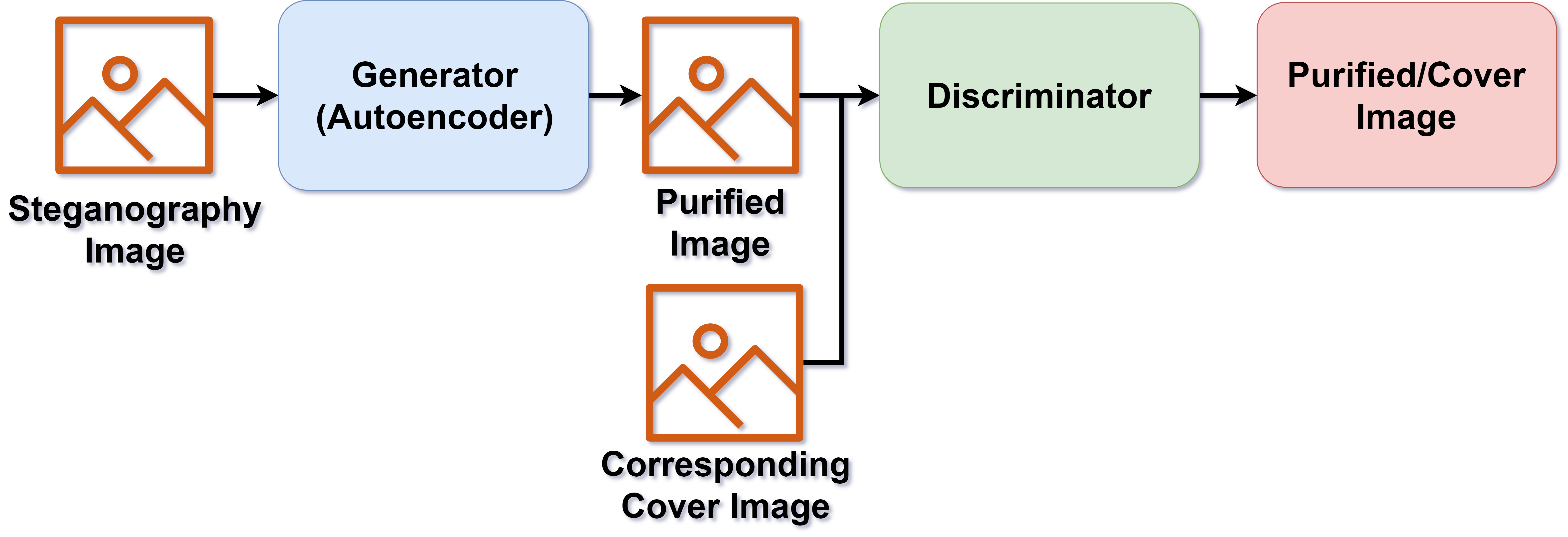}
  \caption{Deep Digital Steganography Purifier (DDSP) Architecture}
  \label{fig:ddsp}
\end{figure}

%% file: figures/autoencoder_arch.tex
\begin{figure}[!t]
  \centering
  \includegraphics[width=0.95\linewidth]{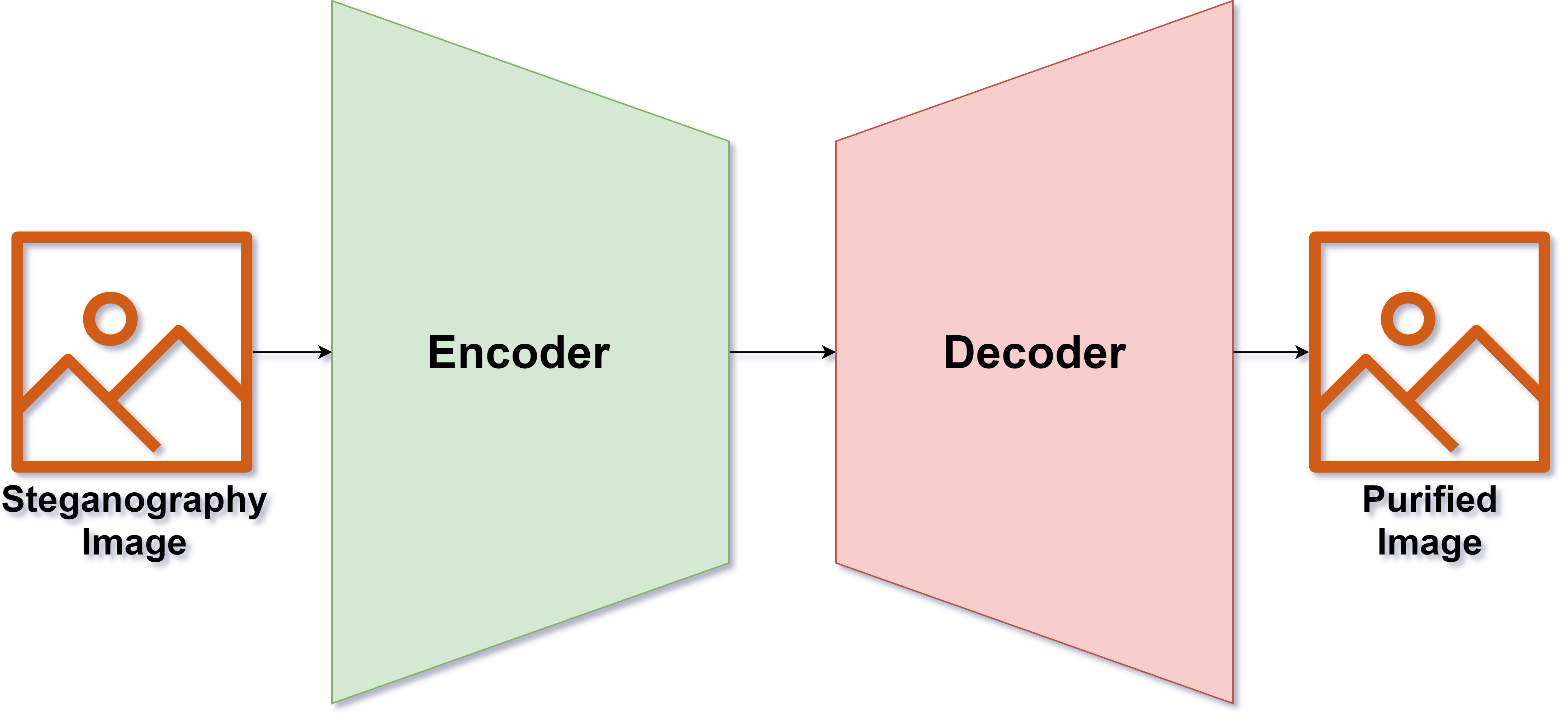}
  \caption{Autoencoder Architecture}
  \label{fig:autoencoder}
\end{figure}

%% file: figures/encoder_arch.tex
\begin{figure}[!t]
  \centering
  \includegraphics[width=0.99\linewidth]{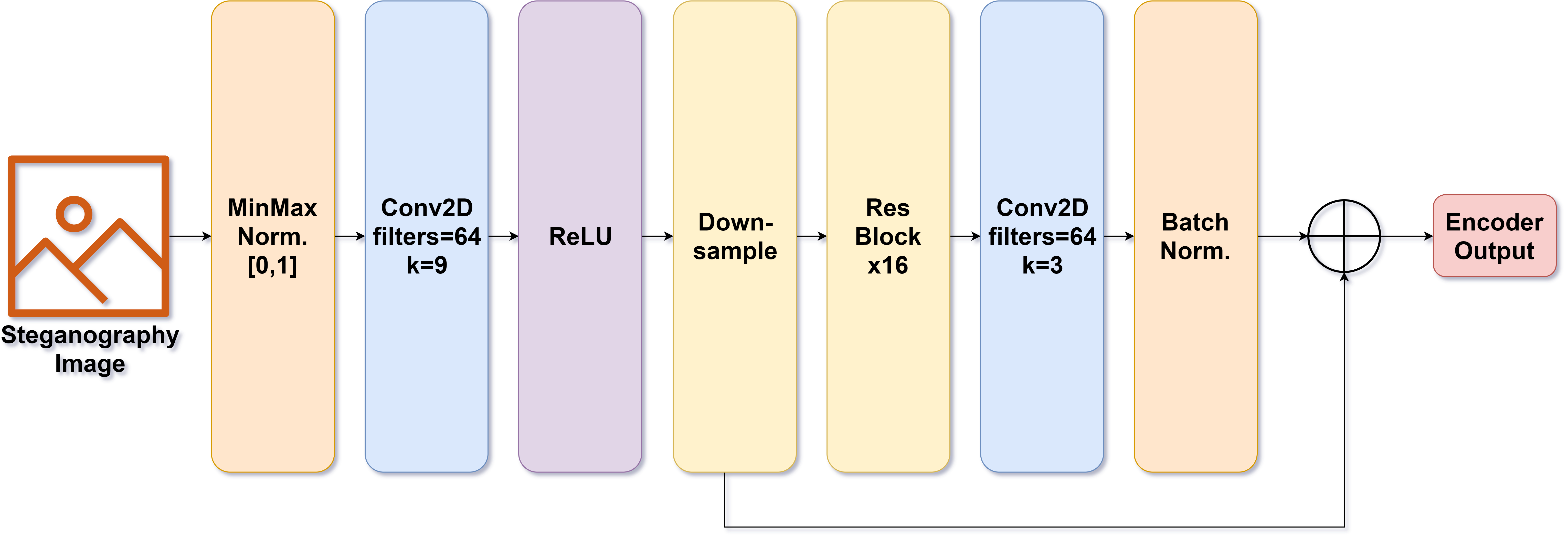}
  \caption{Encoder Network Architecture}
  \label{fig:auto_encoder}
\end{figure}

%% file: sections/dnn.tex
Deep Digital Steganography Purifier (DDSP) consists of a similar architecture to SRGAN \cite{ledig2017photo}. However instead of using a large ResNet, DDSP utilizes a pretrained autoencoder as the generator network in the GAN framework to remove the steganographic content from images without sacrificing image quality. The DDSP model can be seen in Figure \ref{fig:ddsp}. The autoencoder is initially trained using the MSE loss function and then fine tuned using the GAN training framework. This is necessary because autoencoders are trained to optimize MSE which can cause images to have slightly lower quality than the original image due to the reasons discussed in Section \ref{sec:srgan}. More detailed descriptions of the architectures are discussed in the following sections.

\subsection{Autoencoder Architecture}
The residual autoencoder, seen in Figure \ref{fig:autoencoder}, consists of encoder and decoder networks. The encoder learns to reduce the size of the image while maintaining as much information necessary. The decoder then learns how to optimally scale the image to its original size while having removed the steganographic content.

\subsubsection{Encoder Architecture}
The encoder, seen in Figure \ref{fig:auto_encoder}, takes an image with steganography as its input. The input image is then normalized using Min-Max normalization, which scales pixel values to the range of $[0, 1]$. After normalization, the image is fed into a 2-D convolutional layer with a kernel size of 9x9 and 64 filters, followed by a ReLU activation \cite{nair2010rectified}. The output of the convolutional layer is then fed into a downsample block, seen in Figure \ref{fig:downsampler}. The downsample block consists of two 2-D convolutional layers with the first one having a stride of 2 which causes the image to be downsampled by a factor of 2. The use of a convolutional layer to downsample the image is important because it allows the model to learn the near optimal approach to downsample the image while maintaining high image quality. The downsampled output is then passed into 16 serial residual blocks. The residual block architecture can be seen in Figure \ref{fig:res_block}. Following the residual blocks, a final 2-D convolutional layer with batch normalization is added with the original output of the downsample block in a residual manner to form the encoded output.

\subsubsection{Decoder Architecture}
The decoder, seen in Figure \ref{fig:auto_decoder}, takes the output of the encoder as its input. The input is then upsampled using nearest interpolation with a factor of 2. This will increase the shape of the encoder's output back to the size of the original input image. The upsampled image is then fed into a 2-D convolutional layer with a kernel size of 3x3 and 256 filters followed by a ReLU activation. This is then fed into another 2-D convolutional layer with a kernel size of 9x9, and 1 filter followed by a Tanh activation. Since the output of the Tanh activation function is in the range $[-1, 1]$, the output is denormalized to scale the pixel values back to the range $[0, 255]$. This output represents the purified image.

\input{figures/downsample_block_arch.tex}
\input{figures/resblock_arch.tex}

\input{figures/srgan_arch.tex}
\input{figures/decoder_arch.tex}

\subsubsection{Autoencoder Training}
To train the autoencoder, steganographic images are used as the input to the encoder. The encoder creates the encoded image which is fed into the decoder which decodes the image back to its original size. The decoded image is then compared to its corresponding cover image counterpart using the MSE loss function. The autoencoder was trained using early stopping and was optimized using the Adam optimizer \cite{kingma2014adam} with a learning rate $\alpha=10^{-3}$, $\beta_{1}=0.5$, and $\beta_{2}=0.9$.

\subsection{GAN Training}
Similar to the SRGAN training process, we use the pretrained model to initialize the generator network. As seen in Figure \ref{fig:gan_dis}, the discriminator is similar to the SRGAN's discriminator with the exception of DDSP's discriminator blocks, seen in Figure \ref{fig:discrim_block}, which contains the number of convolutional filters per layer in decreasing order to significantly reduce the number of model parameters which decreases training time.

To train the DDSP model, the GAN is trained by having the generator produce purified images. These purified images along with original cover images are passed to the discriminator which is then optimized to distinguish between purified and cover images. The GAN framework was trained for 5 epochs (enough epochs for the model to converge) to fine tune the generator to produce purified images with high frequency detail of the original cover image more accurately.

%% file: figures/downsample_block_arch.tex
\begin{figure}[!t]
  \centering
  \includegraphics[width=0.75\linewidth]{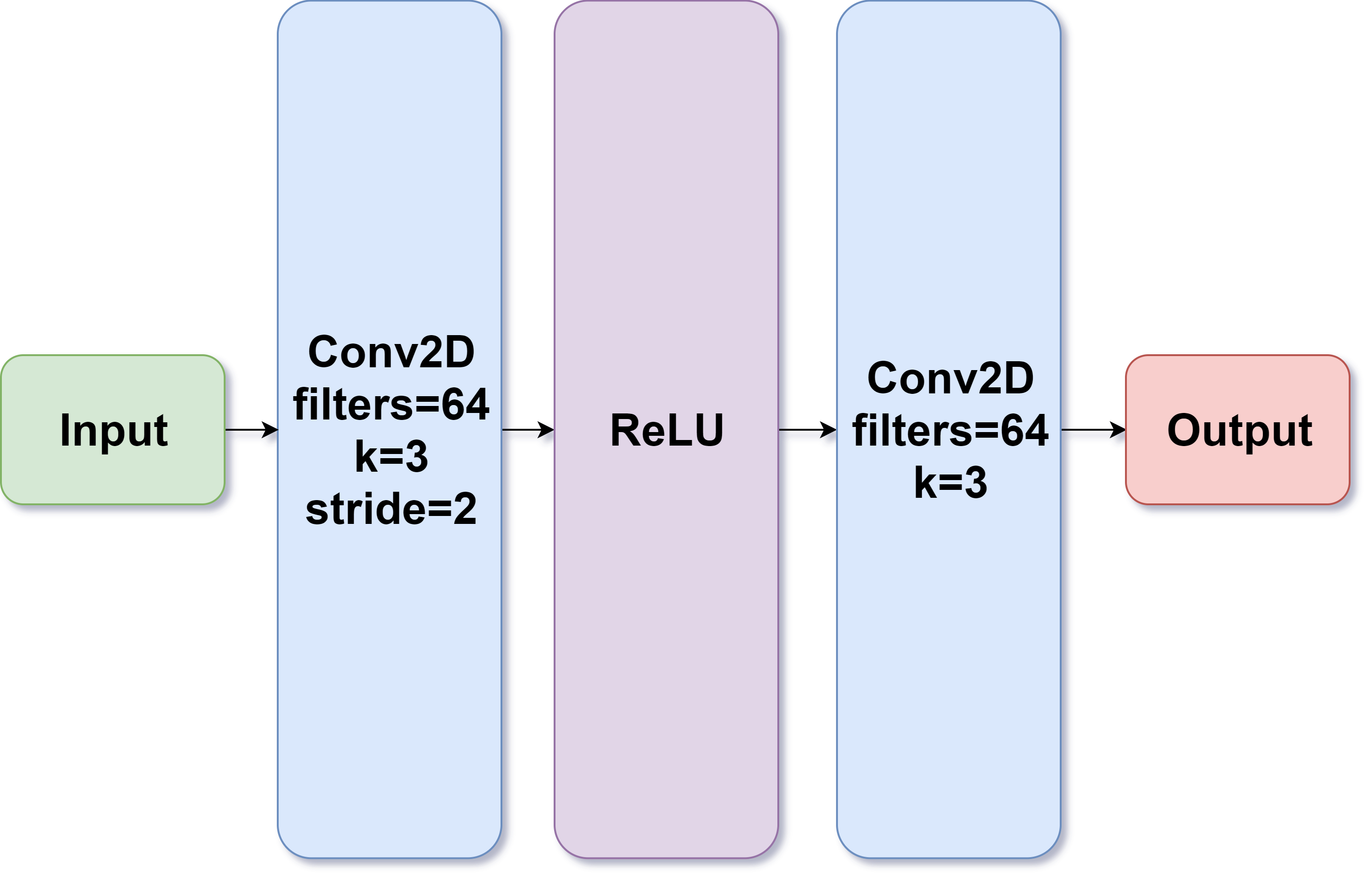}
  \caption{Downsample Block Architecture}
  \label{fig:downsampler}
\end{figure}

%% file: figures/resblock_arch.tex
\begin{figure}[!t]
  \centering
  \includegraphics[width=0.99\linewidth]{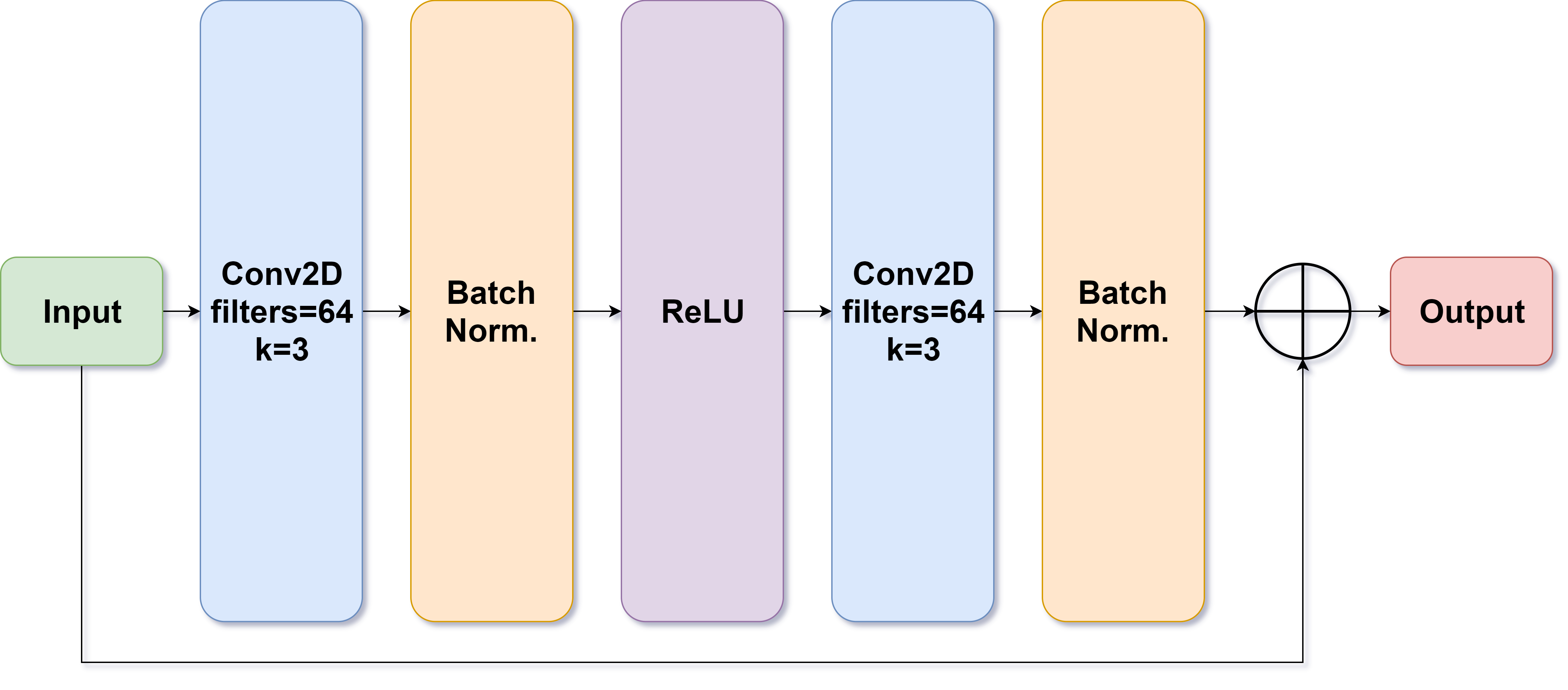}
  \caption{Residual Block Architecture}
  \label{fig:res_block}
\end{figure}

%% file: figures/srgan_arch.tex
\begin{figure*}[!h]
  \centering
  \includegraphics[width=0.99\linewidth]{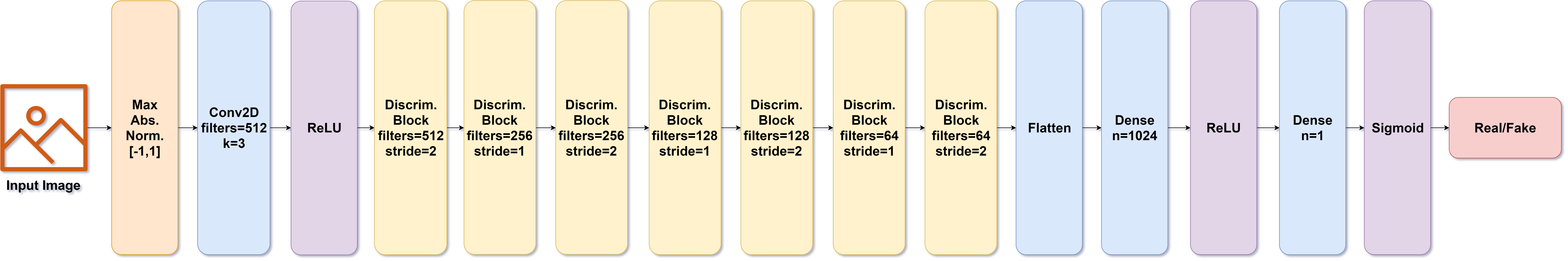}
  \caption{Discriminator Architecture for the DDSP}
  \label{fig:gan_dis}
\end{figure*}

\begin{figure}[!h]
  \centering
  \includegraphics[width=0.75\linewidth]{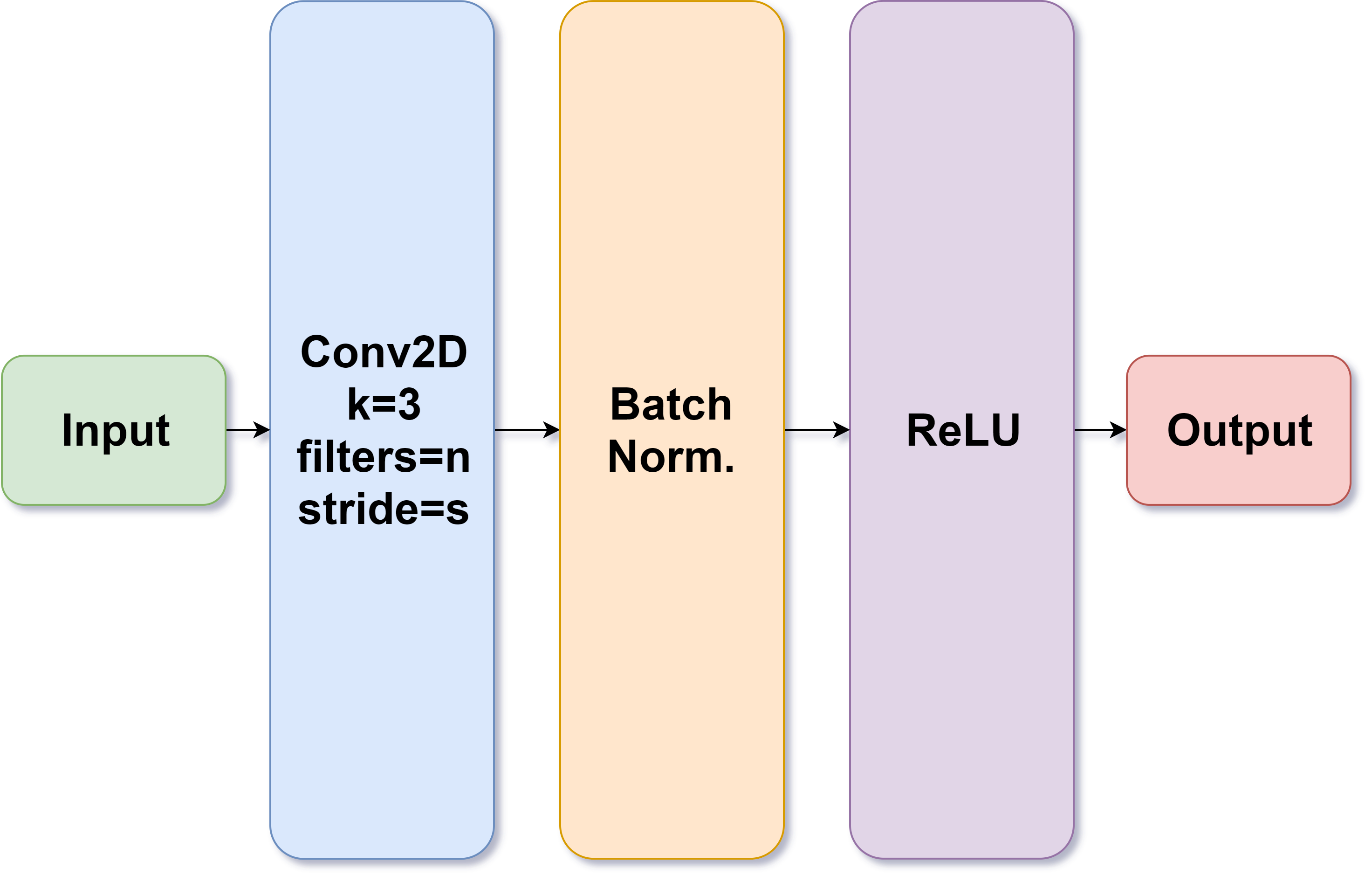}
  \caption{Discriminator Block Architecture}
  \label{fig:discrim_block}
\end{figure}

%% file: figures/decoder_arch.tex
\begin{figure}[!h]
  \centering
  \includegraphics[width=0.99\linewidth]{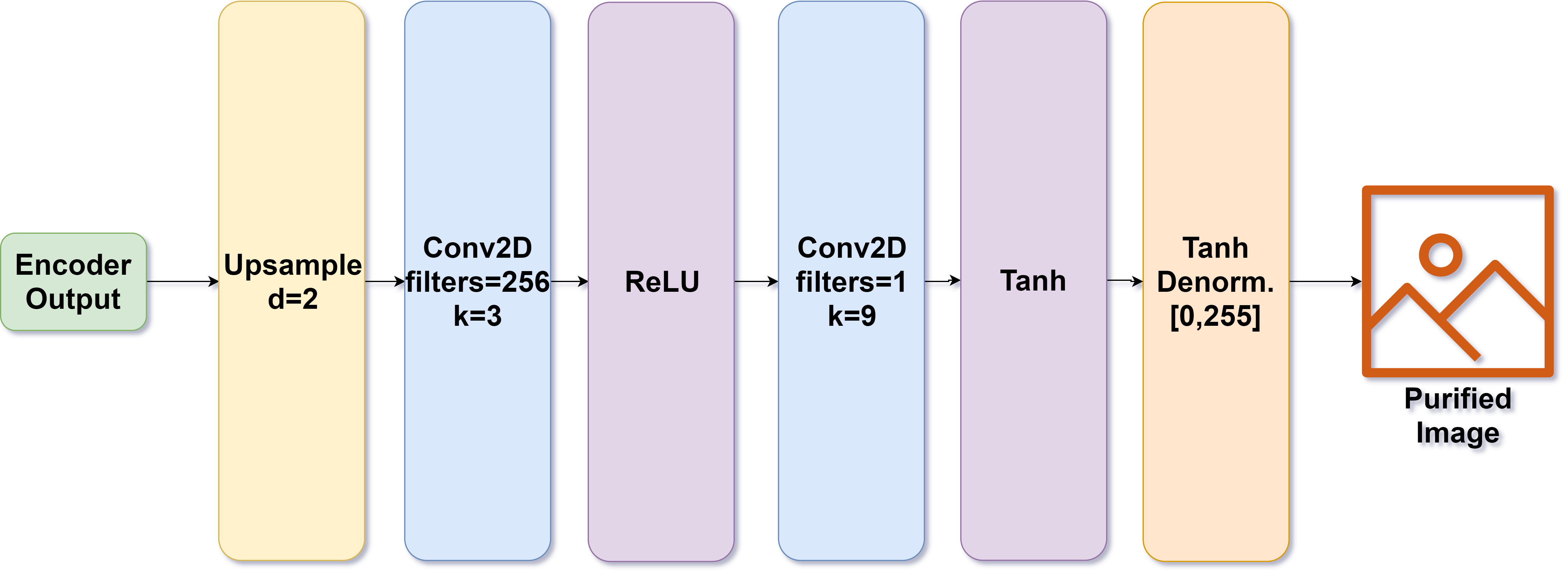}
  \caption{Decoder Network Architecture}
  \label{fig:auto_decoder}
\end{figure}

%% file: sections/results.tex
\input{figures/scrubbed_images.tex}

To assess the performance of our proposed DDSP model in comparison to other steganography removal methods, image resizing and denoising wavelet filters, the testing dataset was used to analyze the image purification quality of the DDSP model. Image quality metrics in combination with visual analysis of image differencing are used to provide further insight to how each method purifies the images. Finally, the DDSP model's generalization abilities are analyzed by testing the transfer learning performance of purifying steganography embedded using different steganography algorithms and file types.

\subsection{Image Purification Quality}
\label{sec:sim}
To compare the quality of the resulting purified images, the following metrics were calculated between the purified images and their corresponding steganographic counterpart images: Mean Squared Error (MSE), Peak Signal-to-Noise Ratio (PSNR), Structural Similarity Index (SSIM) \cite{wang2004image_ssim}, and Universal Quality Index (UQI) \cite{wang2002universal_uqi}. The MSE and PSNR metrics are point-wise measurements of error while the SSIM and UQI metrics were developed to specifically assess image quality. To provide a quantitative measurement of the model's distortion of the pixels to destroy steganographic content, we utilize the bit error ratio (BER) metric, which in our use case can be summarized as the number of bits in the image that have changed after purification, normalized by the total number of bits in the image.

Our proposed DDSP is then baselined against several steganography removal or obfuscation techniques. The first method simply employs bicubic interpolation to downsize an image by a scale factor of 2 and then resize the image back to its original size. As seen in Figure \ref{fig:scrubbed_imgs}(\subref{fig:bicubic_scrub}), the purified image using bicubic interpolation is blurry and does not perform well with respect to maintaining high perceptual image quality. The next baseline method consists of denoising filters using Daubechies 1 (db1) wavelets \cite{daubechies1992ten} and BayesShrink thresholding \cite{chang2000adaptive}. An example of the resulting denoised image can be seen in Figure \ref{fig:scrubbed_imgs}(\subref{fig:wavelet_scrub}). It is notable that the wavelet denoising method is more visually capable of maintaining adequate image quality in comparison to the bicubic resizing method. The final baseline method we compare our DDSP model against is using the pretrained autoencoder prior to GAN fine tuning as the purifier. As seen in Figure \ref{fig:scrubbed_imgs}(\subref{fig:autoencoder_scrub}), the autoencoder does a sufficient job in maintaining image quality while purifying the image. Finally, the resulting purified image from the DDSP can be seen in Figure \ref{fig:scrubbed_imgs}(\subref{fig:gan_scrub}). The DDSP and the autoencoder's purified images have the best visual image quality, with the wavelet filtered image having a slightly lower image quality.

Not only does our proposed DDSP maintain very high perceptual image quality, it is quantitatively better image purifier based on image quality metrics. As seen in Table \ref{tbl:results_bossbase}, the images purified using DDSP resulted in the greatest performance with respect to the BER, MSE, PSNR, SSIM and UQI metrics in comparison to all baselined methods. Since our proposed DDSP model resulted in the highest BER at 82\%, it changed the greatest amount of bits in the image, effectively obfuscating the most amount of steganographic content. Even though our proposed DDSP model changed the highest amount of bits within each image, it produces outputs with the highest quality as verified by the PSNR, SSIM, and UQI metrics, indicating it is the paramount method to use for steganography destruction.

\input{figures/image_diff.tex}

\input{tables/bossbase_results.tex}

\subsection{Image Differencing}
To provide additional analysis of the different image purification models, we subtract the original cover image from their corresponding purified images allowing for the visualization of the effects caused by steganography and purification. As seen in Figure \ref{fig:pix_differences}(\subref{fig:cover_stego}), when the cover image and the corresponding steganographic image are differenced, the resulting image contains a lot of noise. This is expected because the steganography algorithm injects payloads as high frequency noise into the images. The differenced bicubic interpolation purified image, seen in Figure \ref{fig:pix_differences}(\subref{fig:cover_resize}), removes the majority of noise from the image. However, as discussed in the previous section, the bicubic interpolation method does not maintain good visual quality as it removes original content from the image. As seen in Figure \ref{fig:pix_differences}(\subref{fig:cover_wavelet}) and Figure \ref{fig:pix_differences}(\subref{fig:cover_ae}), both the denoising wavelet filter and autoencoder purifier do not remove the noise from the image. Instead, they both appear to inject additional noise into the image to obfuscate the steganographic content, making it unusable. This is visually apparent in the noise located in the whitespace near the top building within the image. For both the wavelet filter and autoencoder, this noise is visually increased in comparison to the original steganographic image. Lastly, as seen in Figure \ref{fig:pix_differences}(\subref{fig:cover_gan}), the DDSP model removes the noise from the image instead of injecting additional noise. This is again apparent in the whitespace near the top of the image. In the DDSP's purified image, almost all of the noise has been removed from these areas, effectively learning to optimally remove the steganographic pattern, which we infer makes the DDSP have the highest image quality in comparison to other methods.

\subsection{Transfer Learning}

Transfer learning can be described as using a applying a model's knowledge gained while training on a certain task to a completely different task. To understand the generalization capability of our model, we form experiments involving the purification of images embedded using an unseen steganography algorithm along with an unseen image format. Additionally we test the purification method of audio files embedded with an unseen steganography algorithm.

\subsubsection{Application to LSB Steganography} \label{sec:image-transfer-learning}
To test the generalization of the DDSP model across unseen image steganography algorithms, we recorded the purification performance of the BOSSBase dataset in its original PGM file format embedded with steganographic payloads using LSB steganography \cite{lsb}. The images were embedded with real malicious payloads generated using Metasploit's MSFvenom payload generator \cite{msfvenom}, which is a commonly used exploitation tool. This was done to mimic the realism of an APT hiding malware using image steganography. Without retraining, the LSB steganography images were purified using the various methods. Similar to the results in Section \ref{sec:sim}, the DDSP model removed the greatest amount of steganography while maintaining the highest image quality. These results can be verified quantitatively by looking at Table \ref{tbl:results_lsb}.

\input{tables/lsb_results.tex}

\subsection{Application to Audio Steganography}
To test the generalization of DDSP across different file formats, we additionally recorded performance metrics on audio files embedded with the same malicious payloads detailed in Section \ref{sec:image-transfer-learning}, using the LSB algorithm. The audio files were from the VoxCeleb1 dataset \cite{voxceleb}, which contains over 1000 utterances from over 12000 speakers, however we only utilized their testing dataset. The testing dataset contains 40 speakers, and 4874 utterances. In order to use the DDSP model without retraining for purifying the audio files, the audio files were reshaped from vectors into matrices and then fed into the DDSP model. The output matrices from the DDSP model were then reshaped back to the original vector format to recreate the audio file. After vectorization, a butterworth lowpass filter \cite{butterworth1930theory} and a hanning window filter \cite{essenwanger1986elements} were applied to the audio file to remove the high frequency edge artifacts created when vectorizing the matrices. The models were baselined against a 1-D denoising wavelet filter as well as upsampling the temporal resolution of the audio signal using bicubic interpolation after downsampling by a scale factor of 2. As seen in Table \ref{tbl:results_audio}, the pretrained autoencoder, denoising wavelet filter, and DDSP are all capable of successfully obfuscating the steganography within the audio files without sacrificing the quality of the audio, with respect to the BER, MSE, and PSNR metrics. However, the upsampling using bicubic interpolation method provides worse MSE and PSNR in comparison to the other techniques. This shows that those models are generalized and remove steganographic content in various file types and steganography algorithms. Although the wavelet denoising filter has slightly better metrics than the DDSP and the pretrained autoencoder, we believe that the DDSP model would greatly outperform wavelet filtering if trained to simultaneously remove image and audio steganography and appropriately handle 1-D signals as input.

\input{tables/audio_results.tex}

%% file: figures/scrubbed_images.tex
\begin{figure*}[!t]
\centering
\begin{subfigure}{0.32\linewidth}
\centering
\includegraphics[trim={0 0 0 0},clip,width=\textwidth]{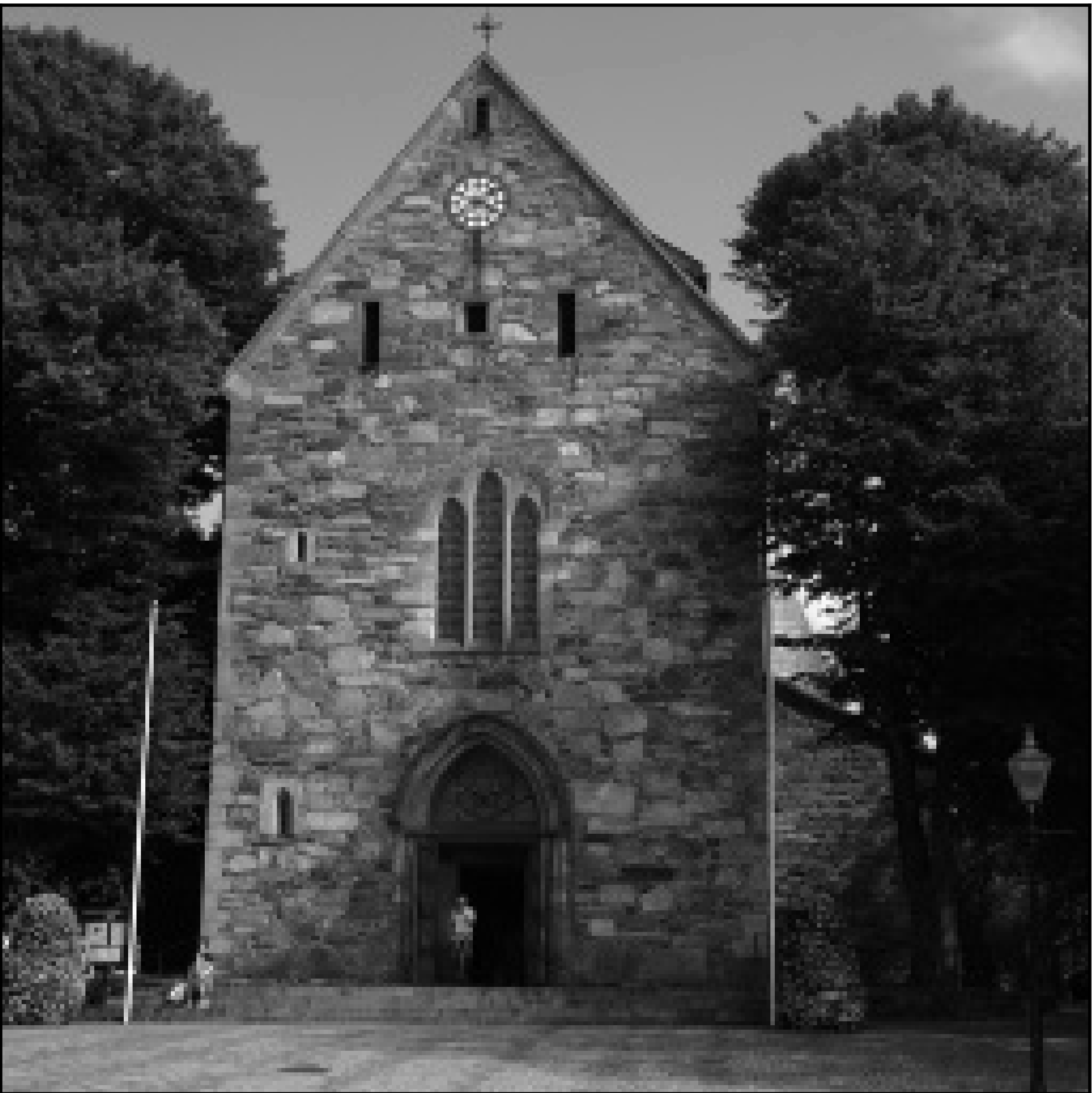}
\caption{Original Image}
\label{fig:cover}
\end{subfigure}
\begin{subfigure}{0.32\linewidth}
\centering
\includegraphics[trim={0 0 0 0},clip,width=\textwidth]{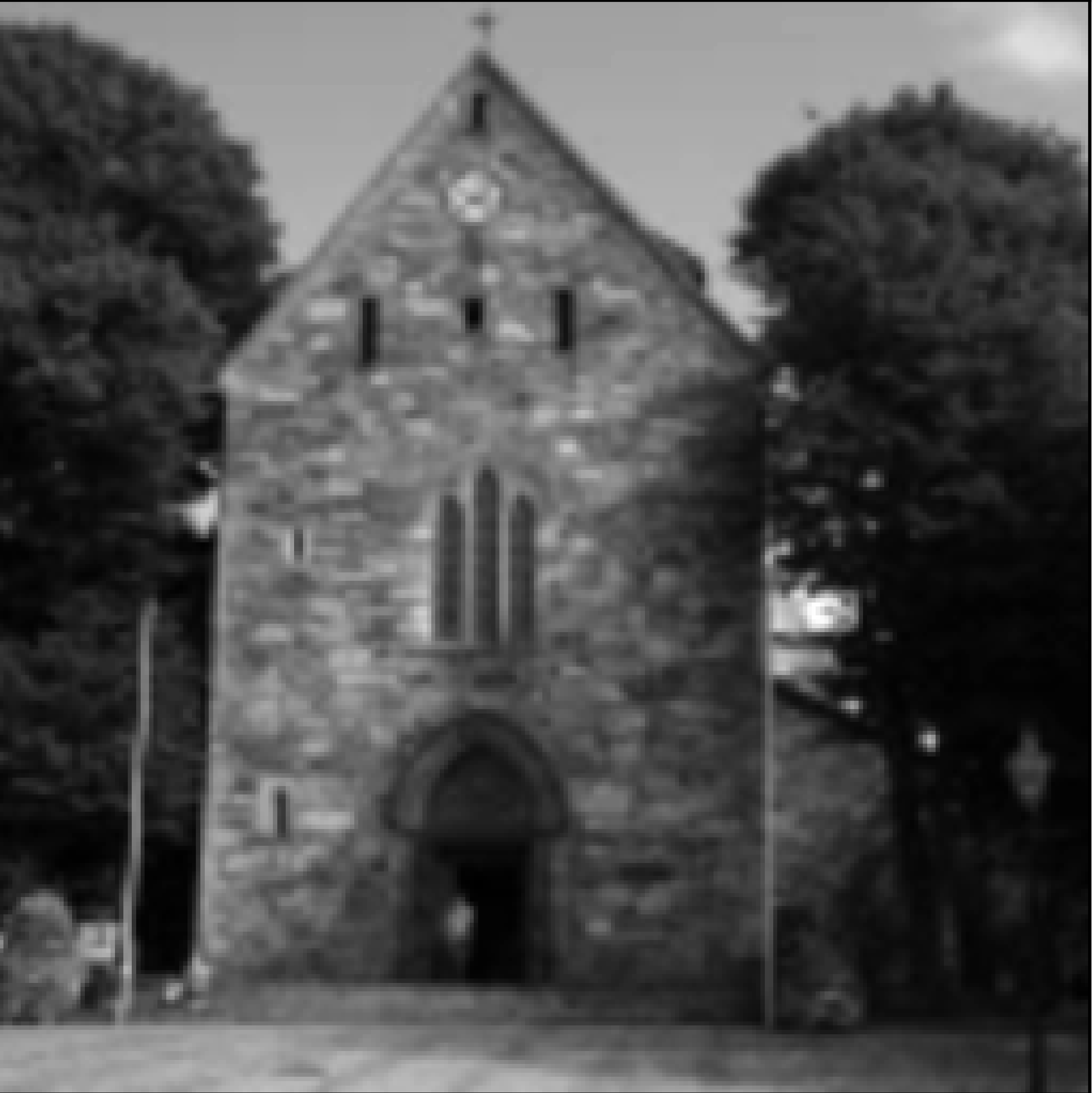}
\caption{Bicubic Interpolation.}
\label{fig:bicubic_scrub}
\end{subfigure}
\begin{subfigure}{0.32\linewidth}
\centering
\includegraphics[trim={0 0 0 0},clip,width=\textwidth]{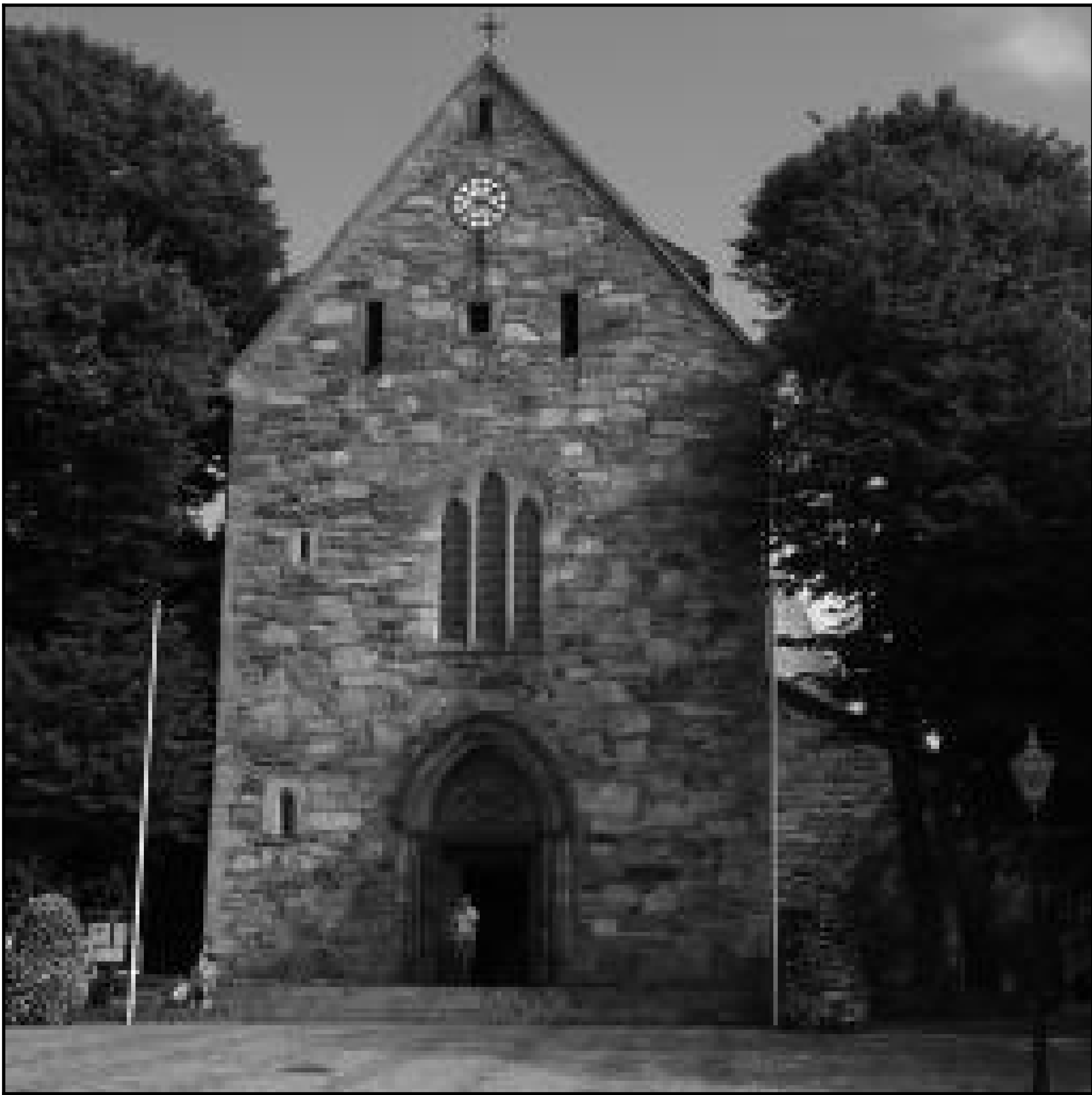}
\caption{Denoising Wavelet Filter}
\label{fig:wavelet_scrub}
\end{subfigure}

\begin{subfigure}{0.32\linewidth}
\centering
\includegraphics[trim={0 0 0 0},clip,width=\textwidth]{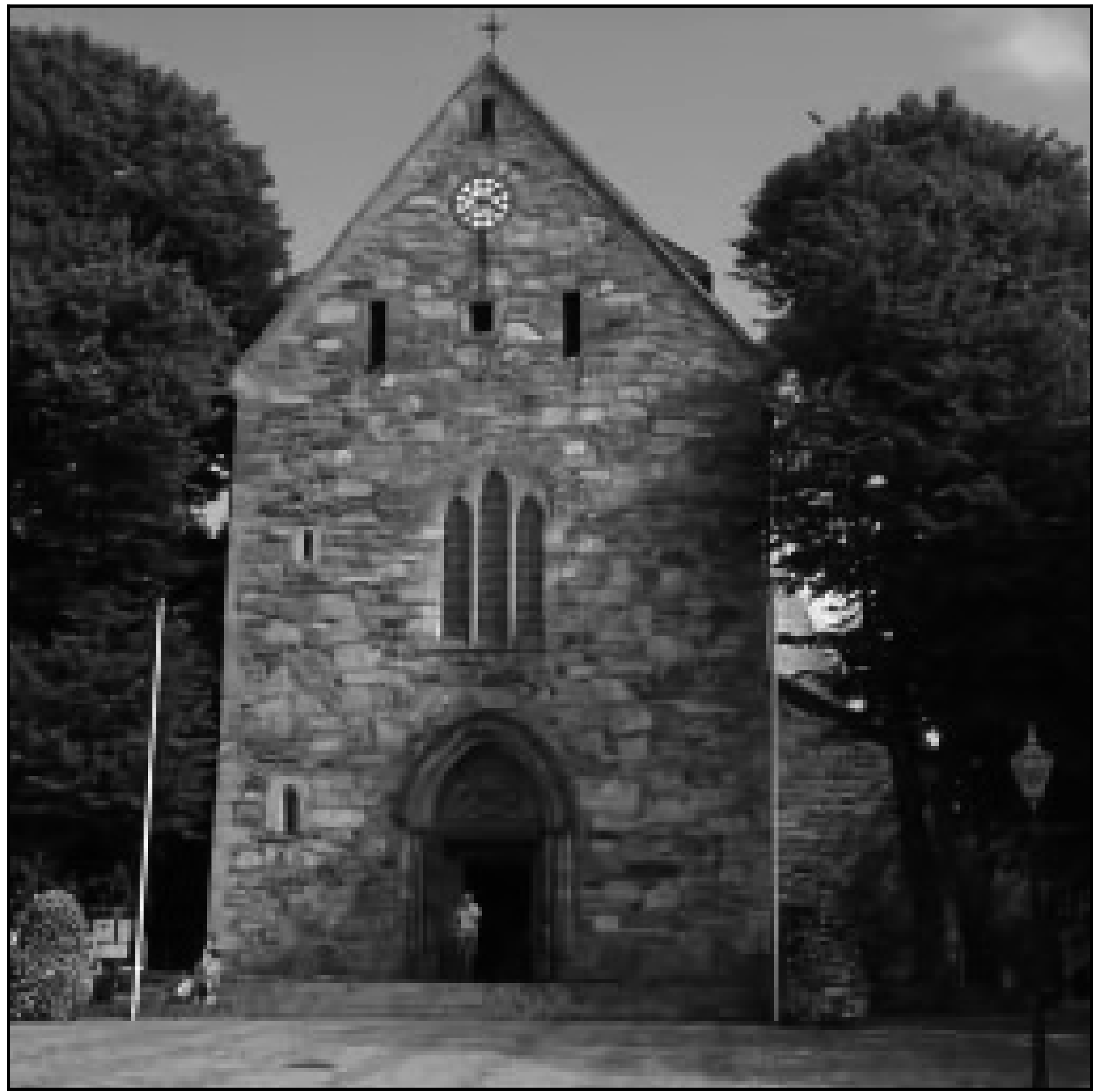}
\caption{Autoencoder}
\label{fig:autoencoder_scrub}
\end{subfigure}
\begin{subfigure}{0.32\linewidth}
\centering
\includegraphics[trim={0 0 0 0},clip,width=\textwidth]{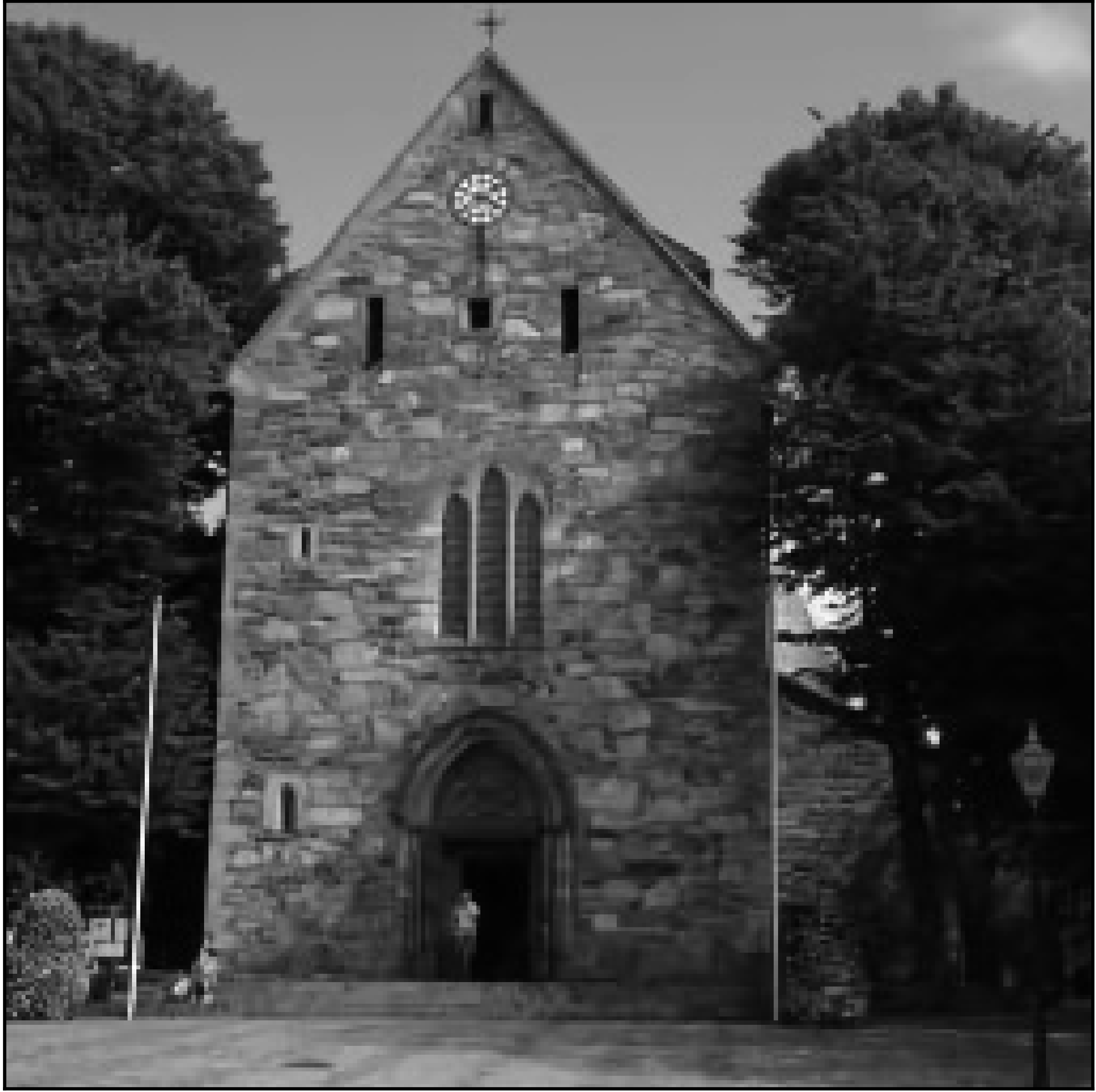}
\caption{DDSP}
\label{fig:gan_scrub}
\end{subfigure}

\caption{Examples of Scrubbed Images Using Various Models}
\label{fig:scrubbed_imgs}
\end{figure*}

%% file: figures/image_diff.tex
\begin{figure*}[!t]
\centering
\begin{subfigure}{0.32\linewidth}
\centering
\includegraphics[trim={0 0 0 0},clip,width=\textwidth]{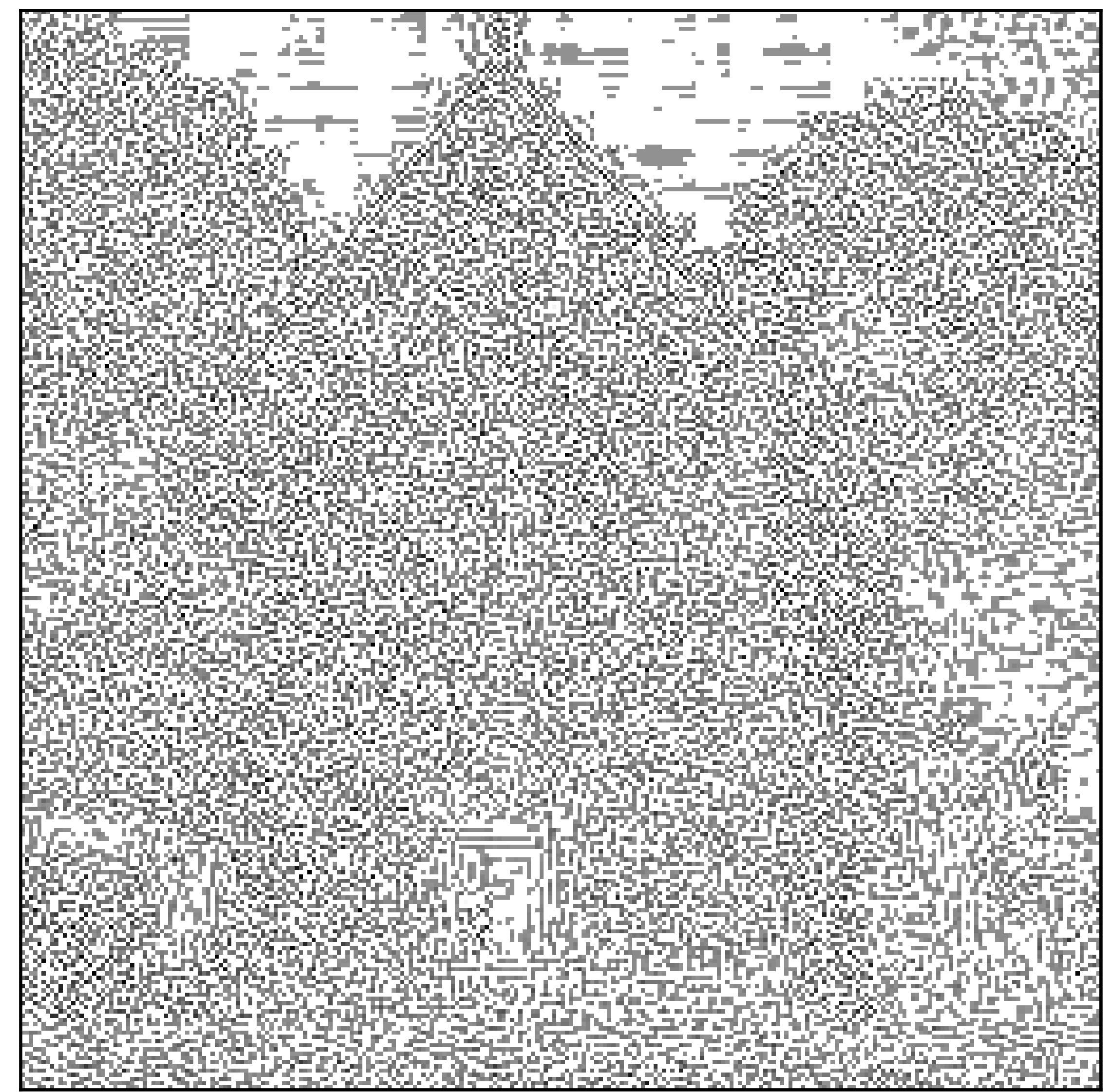}
\caption{Original Steganography}
\label{fig:cover_stego}
\end{subfigure}
\begin{subfigure}{0.32\linewidth}
\centering
\includegraphics[trim={0 0 0 0},clip,width=\textwidth]{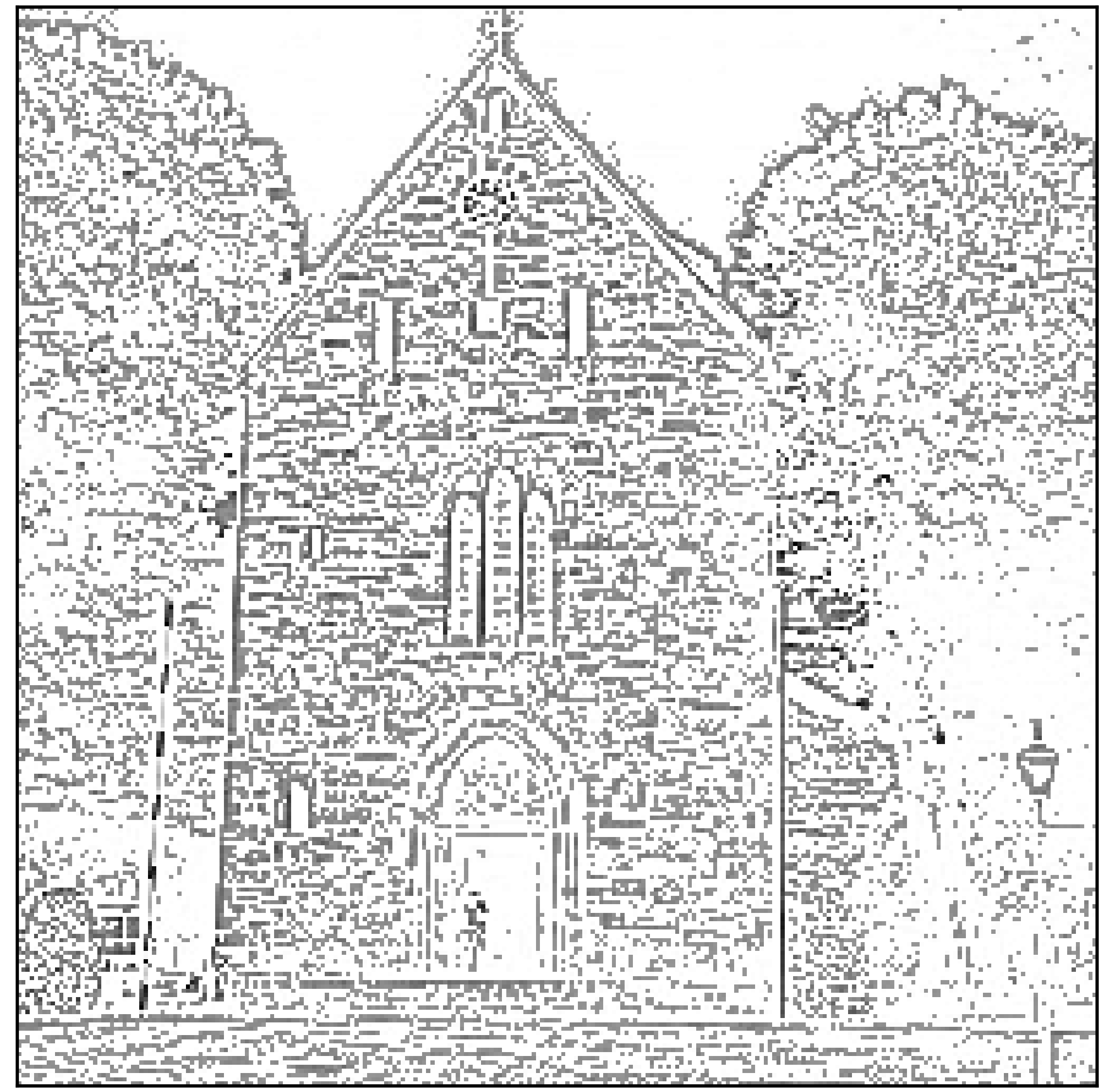}
\caption{Bicubic Interpolation}
\label{fig:cover_resize}
\end{subfigure}
\begin{subfigure}{0.32\linewidth}
\centering
\includegraphics[trim={0 0 0 0},clip,width=\textwidth]{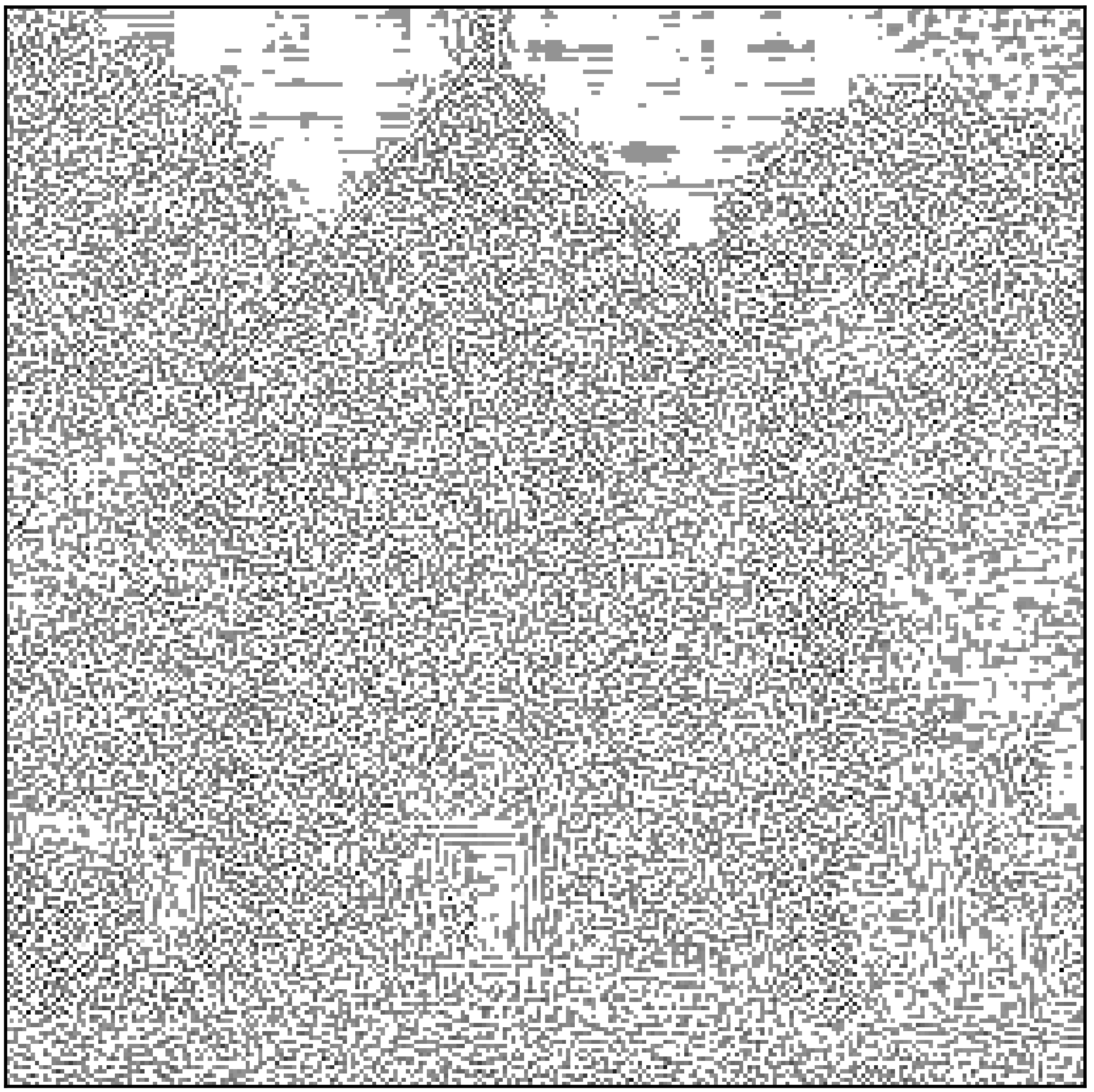}
\caption{Denoising Wavelet Filter}
\label{fig:cover_wavelet}
\end{subfigure}

\begin{subfigure}{0.32\linewidth}
\centering
\includegraphics[trim={0 0 0 0},clip,width=\textwidth]{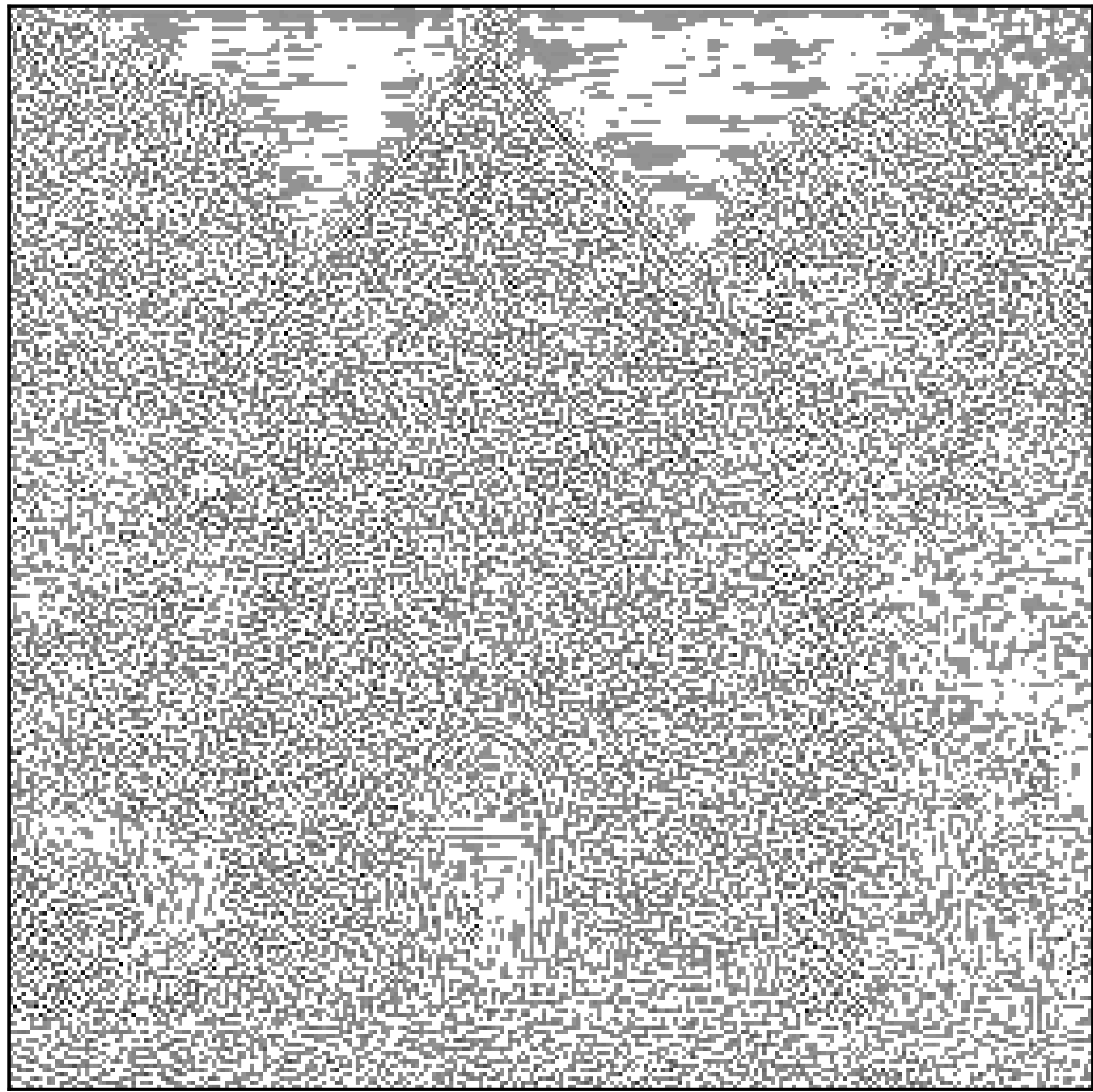}
\caption{Autoencoder}
\label{fig:cover_ae}
\end{subfigure}
\begin{subfigure}{0.32\linewidth}
\centering
\includegraphics[trim={0 0 0 0},clip,width=\textwidth]{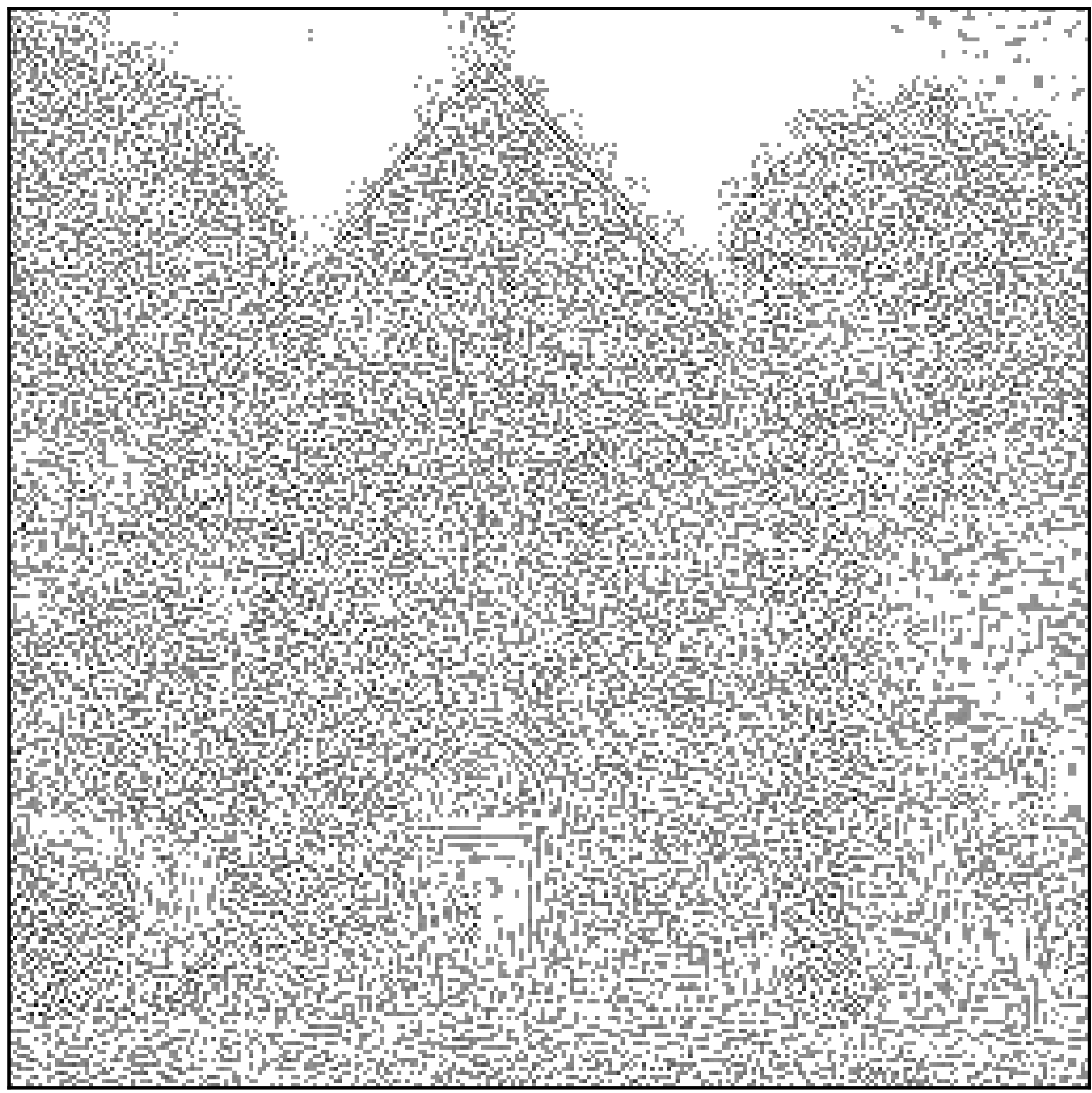}
\caption{DDSP}
\label{fig:cover_gan}
\end{subfigure}

\caption{Examples of Original Image Subtracted from Using Scrubbed Image from Various Models}
\label{fig:pix_differences}
\end{figure*}

%% file: tables/bossbase_results.tex
\begin{table}[!h]
\centering
\caption{Testing Results on the BossBase Dataset}
\label{tbl:results_bossbase}
\resizebox{\linewidth}{!}{%
\begin{tabular}{@{}lrrrrrr@{}}
\toprule
\textbf{Model} & \textbf{BER} & \textbf{MSE} & \textbf{PSNR} & \textbf{SSIM} & \textbf{UQI} \\ \midrule
DDSP & \textbf{0.82} & \textbf{5.27} & \textbf{40.91} & \textbf{0.99} & \textbf{0.99} \\
Autoencoder & 0.78 & 5.97 & 40.37 & 0.98 & \textbf{0.99} \\
Wavelet Filter & 0.52 & 6942.51 & 9.72 & 0.19 & 0.50 \\
Bicubic Inter. & 0.53 & 6767.35 & 9.82 & 0.22 & 0.51  \\ \bottomrule
\end{tabular}%
}
\end{table}

%% file: tables/lsb_results.tex
\begin{table}[!h]
\centering
\caption{Transfer Learning Results on the LSB Dataset}
\label{tbl:results_lsb}
\resizebox{\linewidth}{!}{%
\begin{tabular}{@{}lllllll@{}}
\toprule
\textbf{Model} & \textbf{BER} & \textbf{MSE} & \textbf{PSNR} & \textbf{SSIM} & \textbf{UQI} \\ \midrule
DDSP & \textbf{0.82} & \textbf{5.09} & \textbf{41.05} & \textbf{0.98} & \textbf{0.99} \\
Autoencoder & 0.78 & 5.63 & 40.62 & 0.98 & \textbf{0.99} \\
Wavelet Filter & 0.53 & 6935.08 & 9.72 & 0.19 & 0.50 \\
Bicubic Inter. & 0.53 & 6763.73 & 9.83 & 0.22 & 0.51 \\ \bottomrule
\end{tabular}%
}
\end{table}

%% file: tables/audio_results.tex
\begin{table}[!h]
\centering
\caption{Transfer Learning Results on the VoxCeleb1 Dataset}
\label{tbl:results_audio}
\resizebox{0.75\linewidth}{!}{%
\begin{tabular}{@{}lllllll@{}}
\toprule
\textbf{Model} & \textbf{BER} & \textbf{MSE} & \textbf{PSNR} \\ \midrule
DDSP & \textbf{0.67} & 650.12 & 37.28 \\
Autoencoder & \textbf{0.67} & 650.14 & 37.28 \\
Wavelet Filter & \textbf{0.67} & \textbf{643.94} & \textbf{37.65} \\
Bicubic Inter. & 0.64 & 1157.01 & 35.54 \\ \bottomrule
\end{tabular}%
}
\end{table}

%% file: sections/conclusion.tex
In this paper, we developed a steganography purification model which we term Deep Digital Steganography Purifier (DDSP), which utilizes a Generative Adversarial Network (GAN) which, to the best of our knowledge, removes the highest amount of steganographic content from images while maintaining the highest visual image quality in comparison to other state-of-the-art techniques as verified by visual and quantitative results. In the future, we plan to extend the DDSP model to purify inputs of various types, sizes, and color spaces. Additionally we plan to train the DDSP model on a larger dataset to make the model more robust, thus making it ready to be operationalized for a real steganography purification system.

%% file: sections/acknowledgements.tex
The authors would like to thank the members of our team for their assistance, guidance, and review of our research.

%% file: bare_jrnl.bbl
\begin{thebibliography}{10}
\providecommand{\url}[1]{#1}
\csname url@samestyle\endcsname
\providecommand{\newblock}{\relax}
\providecommand{\bibinfo}[2]{#2}
\providecommand{\BIBentrySTDinterwordspacing}{\spaceskip=0pt\relax}
\providecommand{\BIBentryALTinterwordstretchfactor}{4}
\providecommand{\BIBentryALTinterwordspacing}{\spaceskip=\fontdimen2\font plus
\BIBentryALTinterwordstretchfactor\fontdimen3\font minus
  \fontdimen4\font\relax}
\providecommand{\BIBforeignlanguage}[2]{{%
\expandafter\ifx\csname l@#1\endcsname\relax
\typeout{** WARNING: IEEEtran.bst: No hyphenation pattern has been}%
\typeout{** loaded for the language `#1'. Using the pattern for}%
\typeout{** the default language instead.}%
\else
\language=\csname l@#1\endcsname
\fi
#2}}
\providecommand{\BIBdecl}{\relax}
\BIBdecl

\bibitem{cox2007digital}
I.~Cox, M.~Miller, J.~Bloom, J.~Fridrich, and T.~Kalker, \emph{Digital
  watermarking and steganography}.\hskip 1em plus 0.5em minus 0.4em\relax
  Morgan kaufmann, 2007.

\bibitem{apt37}
\BIBentryALTinterwordspacing
``Apt37.'' [Online]. Available: \url{https://attack.mitre.org/groups/ G0067/}
\BIBentrySTDinterwordspacing

\bibitem{bender1996techniques}
W.~Bender, D.~Gruhl, N.~Morimoto, and A.~Lu, ``Techniques for data hiding,''
  \emph{IBM systems journal}, vol.~35, no. 3.4, pp. 313--336, 1996.

\bibitem{goodfellow2014generative}
I.~Goodfellow, J.~Pouget-Abadie, M.~Mirza, B.~Xu, D.~Warde-Farley, S.~Ozair,
  A.~Courville, and Y.~Bengio, ``Generative adversarial nets,'' in
  \emph{Advances in neural information processing systems}, 2014, pp.
  2672--2680.

\bibitem{westfeld1999attacks}
A.~Westfeld and A.~Pfitzmann, ``Attacks on steganographic systems,'' in
  \emph{International workshop on information hiding}.\hskip 1em plus 0.5em
  minus 0.4em\relax Springer, 1999, pp. 61--76.

\bibitem{jung2019pixelsteganalysis}
D.~Jung, H.~Bae, H.-S. Choi, and S.~Yoon, ``Pixelsteganalysis: Destroying
  hidden information with a low degree of visual degradation,'' \emph{arXiv
  preprint arXiv:1902.11113}, 2019.

\bibitem{salimans2017pixelcnn++}
T.~Salimans, A.~Karpathy, X.~Chen, and D.~P. Kingma, ``Pixelcnn++: Improving
  the pixelcnn with discretized logistic mixture likelihood and other
  modifications,'' \emph{arXiv preprint arXiv:1701.05517}, 2017.

\bibitem{baluja2017hiding}
S.~Baluja, ``Hiding images in plain sight: Deep steganography,'' in
  \emph{Advances in Neural Information Processing Systems}, 2017, pp.
  2069--2079.

\bibitem{zhang2019invisible}
R.~Zhang, S.~Dong, and J.~Liu, ``Invisible steganography via generative
  adversarial networks,'' \emph{Multimedia Tools and Applications}, vol.~78,
  no.~7, pp. 8559--8575, 2019.

\bibitem{amritha2019anti}
P.~Amritha, M.~Sethumadhavan, R.~Krishnan, and S.~K. Pal, ``Anti-forensic
  approach to remove stego content from images and videos,'' \emph{Journal of
  Cyber Security and Mobility}, vol.~8, no.~3, pp. 295--320, 2019.

\bibitem{amritha2016removal}
P.~Amritha, M.~Sethumadhavan, and R.~Krishnan, ``On the removal of
  steganographic content from images,'' \emph{Defence Science Journal},
  vol.~66, no.~6, pp. 574--581, 2016.

\bibitem{ameen2013optimal}
S.~Y. Ameen and M.~R. Al-Badrany, ``Optimal image steganography content
  destruction techniques,'' in \emph{International Conference on Systems,
  Control, Signal Processing and Informatics}, 2013, pp. 453--457.

\bibitem{sharp2013novel}
A.~Sharp, Q.~Qi, Y.~Yang, D.~Peng, and H.~Sharif, ``A novel active warden
  steganographic attack for next-generation steganography,'' in \emph{2013 9th
  International Wireless Communications and Mobile Computing Conference
  (IWCMC)}.\hskip 1em plus 0.5em minus 0.4em\relax IEEE, 2013, pp. 1138--1143.

\bibitem{al2006destroying}
F.~Al-Naima, S.~Y. Ameen, and A.~F. Al-Saad, ``Destroying steganography content
  in image files,'' in \emph{The 5th International Symposium on Communication
  Systems, Networks and DSP (CSNDSP’06), Greece}, 2006.

\bibitem{shrestha2011general}
P.~L. Shrestha, M.~Hempel, T.~Ma, D.~Peng, and H.~Sharif, ``A general attack
  method for steganography removal using pseudo-cfa re-interpolation,'' in
  \emph{2011 International Conference for Internet Technology and Secured
  Transactions}.\hskip 1em plus 0.5em minus 0.4em\relax IEEE, 2011, pp.
  454--459.

\bibitem{amritha2016active}
P.~Amritha, K.~Induja, and K.~Rajeev, ``Active warden attack on steganography
  using prewitt filter,'' in \emph{Proceedings of the International Conference
  on Soft Computing Systems}.\hskip 1em plus 0.5em minus 0.4em\relax Springer,
  2016, pp. 591--599.

\bibitem{smith2007denoising}
C.~B. Smith and S.~S. Agaian, ``Denoising and the active warden,'' in
  \emph{2007 IEEE International Conference on Systems, Man and
  Cybernetics}.\hskip 1em plus 0.5em minus 0.4em\relax IEEE, 2007, pp.
  3317--3322.

\bibitem{ledig2017photo}
C.~Ledig, L.~Theis, F.~Husz{\'a}r, J.~Caballero, A.~Cunningham, A.~Acosta,
  A.~Aitken, A.~Tejani, J.~Totz, Z.~Wang \emph{et~al.}, ``Photo-realistic
  single image super-resolution using a generative adversarial network,'' in
  \emph{Proceedings of the IEEE conference on computer vision and pattern
  recognition}, 2017, pp. 4681--4690.

\bibitem{he2016deep}
K.~He, X.~Zhang, S.~Ren, and J.~Sun, ``Deep residual learning for image
  recognition,'' in \emph{Proceedings of the IEEE conference on computer vision
  and pattern recognition}, 2016, pp. 770--778.

\bibitem{bossbase}
P.~Bas, T.~Filler, and T.~Pevn{\`y}, ``"break our steganographic system": the
  ins and outs of organizing boss,'' in \emph{International workshop on
  information hiding}.\hskip 1em plus 0.5em minus 0.4em\relax Springer, 2011,
  pp. 59--70.

\bibitem{hugo}
T.~Pevn{\`y}, T.~Filler, and P.~Bas, ``Using high-dimensional image models to
  perform highly undetectable steganography,'' in \emph{International Workshop
  on Information Hiding}.\hskip 1em plus 0.5em minus 0.4em\relax Springer,
  2010, pp. 161--177.

\bibitem{hill}
B.~Li, M.~Wang, J.~Huang, and X.~Li, ``A new cost function for spatial image
  steganography,'' in \emph{2014 IEEE International Conference on Image
  Processing (ICIP)}.\hskip 1em plus 0.5em minus 0.4em\relax IEEE, 2014, pp.
  4206--4210.

\bibitem{suniward}
V.~Holub, J.~Fridrich, and T.~Denemark, ``Universal distortion function for
  steganography in an arbitrary domain,'' \emph{EURASIP Journal on Information
  Security}, vol. 2014, no.~1, p.~1, 2014.

\bibitem{wow}
V.~Holub and J.~Fridrich, ``Designing steganographic distortion using
  directional filters,'' in \emph{2012 IEEE International workshop on
  information forensics and security (WIFS)}.\hskip 1em plus 0.5em minus
  0.4em\relax IEEE, 2012, pp. 234--239.

\bibitem{stego_code}
\BIBentryALTinterwordspacing
J.~Fridrich, ``Steganographic algorithms.'' [Online]. Available:
  \url{http://dde.binghamton.edu/download/stego\_algorithms/}
\BIBentrySTDinterwordspacing

\bibitem{nair2010rectified}
V.~Nair and G.~E. Hinton, ``Rectified linear units improve restricted boltzmann
  machines,'' in \emph{Proceedings of the 27th international conference on
  machine learning (ICML-10)}, 2010, pp. 807--814.

\bibitem{kingma2014adam}
D.~P. Kingma and J.~Ba, ``Adam: A method for stochastic optimization,''
  \emph{arXiv preprint arXiv:1412.6980}, 2014.

\bibitem{wang2004image_ssim}
Z.~Wang, A.~C. Bovik, H.~R. Sheikh, E.~P. Simoncelli \emph{et~al.}, ``Image
  quality assessment: from error visibility to structural similarity,''
  \emph{IEEE transactions on image processing}, vol.~13, no.~4, pp. 600--612,
  2004.

\bibitem{wang2002universal_uqi}
Z.~Wang and A.~C. Bovik, ``A universal image quality index,'' \emph{IEEE signal
  processing letters}, vol.~9, no.~3, pp. 81--84, 2002.

\bibitem{daubechies1992ten}
I.~Daubechies, \emph{Ten lectures on wavelets}.\hskip 1em plus 0.5em minus
  0.4em\relax Siam, 1992, vol.~61.

\bibitem{chang2000adaptive}
S.~G. Chang, B.~Yu, and M.~Vetterli, ``Adaptive wavelet thresholding for image
  denoising and compression,'' \emph{IEEE transactions on image processing},
  vol.~9, no.~9, pp. 1532--1546, 2000.

\bibitem{lsb}
P.~Wayner, \emph{Disappearing cryptography: information hiding: steganography
  and watermarking}.\hskip 1em plus 0.5em minus 0.4em\relax Morgan Kaufmann,
  2009.

\bibitem{msfvenom}
\BIBentryALTinterwordspacing
``Msfvenom.'' [Online]. Available:
  \url{https://www.offensive-security.com/metasploit-unleashed/msfvenom/}
\BIBentrySTDinterwordspacing

\bibitem{voxceleb}
A.~Nagrani, J.~S. Chung, and A.~Zisserman, ``Voxceleb: a large-scale speaker
  identification dataset,'' in \emph{INTERSPEECH}, 2017.

\bibitem{butterworth1930theory}
S.~Butterworth \emph{et~al.}, ``On the theory of filter amplifiers,''
  \emph{Wireless Engineer}, vol.~7, no.~6, pp. 536--541, 1930.

\bibitem{essenwanger1986elements}
O.~M. Essenwanger, ``Elements of statistical analysis,'' 1986.

\end{thebibliography}
